\documentclass[english,11pt,a4paper,openany]{IEEEtran}

\usepackage[style=ieee,backend=biber,sorting=ynt]{biblatex}   
\usepackage{url}
\usepackage{graphicx}  
\usepackage{float}  
	
\begin{document}
		
		\title{Performance best practices using Java AWS lambda\\ \small{Measuring enhancements of different optimization techniques combining java with AWS lambda}}
		
		\author{
			Juan Mera Menéndez,
			Martin Bartlett
			}
		
		\markboth{DXC Technology, research. 01, September 2023}
		{Shell \MakeLowercase{\textit{et al.}}: Bare Demo of IEEEtran.cls for IEEE Journals}
		
		\maketitle
		
		\begin{abstract} 
			Despite its already widespread popularity, it continues to gain adoption. More and more developers and architects continue to adopt and apply the FaaS (Function as a Service) model in cloud solutions. The most extensively used FaaS service is AWS Lambda, provided by Amazon Web Services. Moreover, despite the new trends in programming languages, Java still maintains a significant share of usage.
            The main problem that arises when using these two technologies together is widely known: significant latencies and the dreaded cold start. However, it is possible to greatly mitigate this problem without dedicating too much effort.
            In this article, various techniques, strategies and approaches will be studied with the aim of reducing the cold start and significantly improving the performance of Lambda functions with Java. Starting from a system that involves AWS lambda, java, DynamoDB and Api Gateway. Each approach will be tested independently, analyzing its impact through load tests. Subsequently, they will be tested in combination in an effort to achieve the greatest possible performance improvement.  
		\end{abstract}
		
		\begin{IEEEkeywords}
			AWS Lambda, AWS, Amazon Web Services, Java, performance, cold start, FaaS, Function as a Service, Serverless, GraalVM, Snapstart, JAVA\_TOOL\_OPTIONS 
		\end{IEEEkeywords}
  
        \section{Introduction} 
		
		\IEEEPARstart{F}{ollowing} the trend of recent years, the serverless approach \parencite{Hassansurvey} continues to gain popularity due to its scalability, flexibility, agility, and other well-known benefits. According to the survey conducted by Datadog in August 2023 \parencite{Datdog2023}, one of the most popularly used services within this architectural model is AWS Lambda (provided by Amazon Web Services) and over a 10\% of Lambda functions are written in Java. \parencite{StackOverflow2022} Furthermore Java is the primary language for 30\% of professional developers. \\
        Being the performance one of the main criteria when comparing solutions, the survey shows that cold start of java lambdas are roughly twice as long as the cold starts of other languages like Node.js and Python \parencite{Jackson, Hosseini2019}(which are the most used languages with AWS lambda). Indeed, it becomes evident why performance is the first aspect brought to the table when discussing the combination of Java with AWS Lambda. And rightfully so, if it's not taken into account when building the system, it's common to encounter time-out issues in API Gateway due to excessively high latency of the functions. Furthermore, another noteworthy benefit of improving the performance of Lambda functions is the reduction in their cost. This is because the cost model for this service is based on computing time. By reducing this time, the cost is also reduced. \\
        Thus, the primary goal of the study is to optimize the way Java code is executed on AWS Lambda and provide a set of best practices that assist this approach in being competitive in terms of performance. To achieve that goal, a series of techniques, approaches or strategies are proposed with the aim of improving function performance and mitigating cold starts:
        \begin{itemize}
			\item Profile the function to see which memory configuration is better suited in terms of performance.
			\item Use Snapstart to reduce the cold-start latency.
			\item Choose the Arm64 Lambda architecture against the default x86\_64 architecture.
            \item Use the AWS SDK v2 for Java.
            \item Pre-compile the source code with GraalVM to avoid initialize all classes in runtime.
            \item Java Lambda function customization settings: JAVA\_TOOL\_OPTIONS environment variable.
            \item Other good practices
		\end{itemize}
        The improvement caused by each technique will be measured independently since some of them are not compatible with each other. The advantages and drawbacks of the techniques will be discussed as well. For the implementation of each technique or approach, we start with an initial, unoptimized system, which will be explained in the following section. The performance of this system has also been measured and will be used to calculate the improvements experienced relative to it. After this, the techniques will be combined, and the result will be measured.
		\begin{figure}[h]
			\centering
			\includegraphics[width=0.7\linewidth]{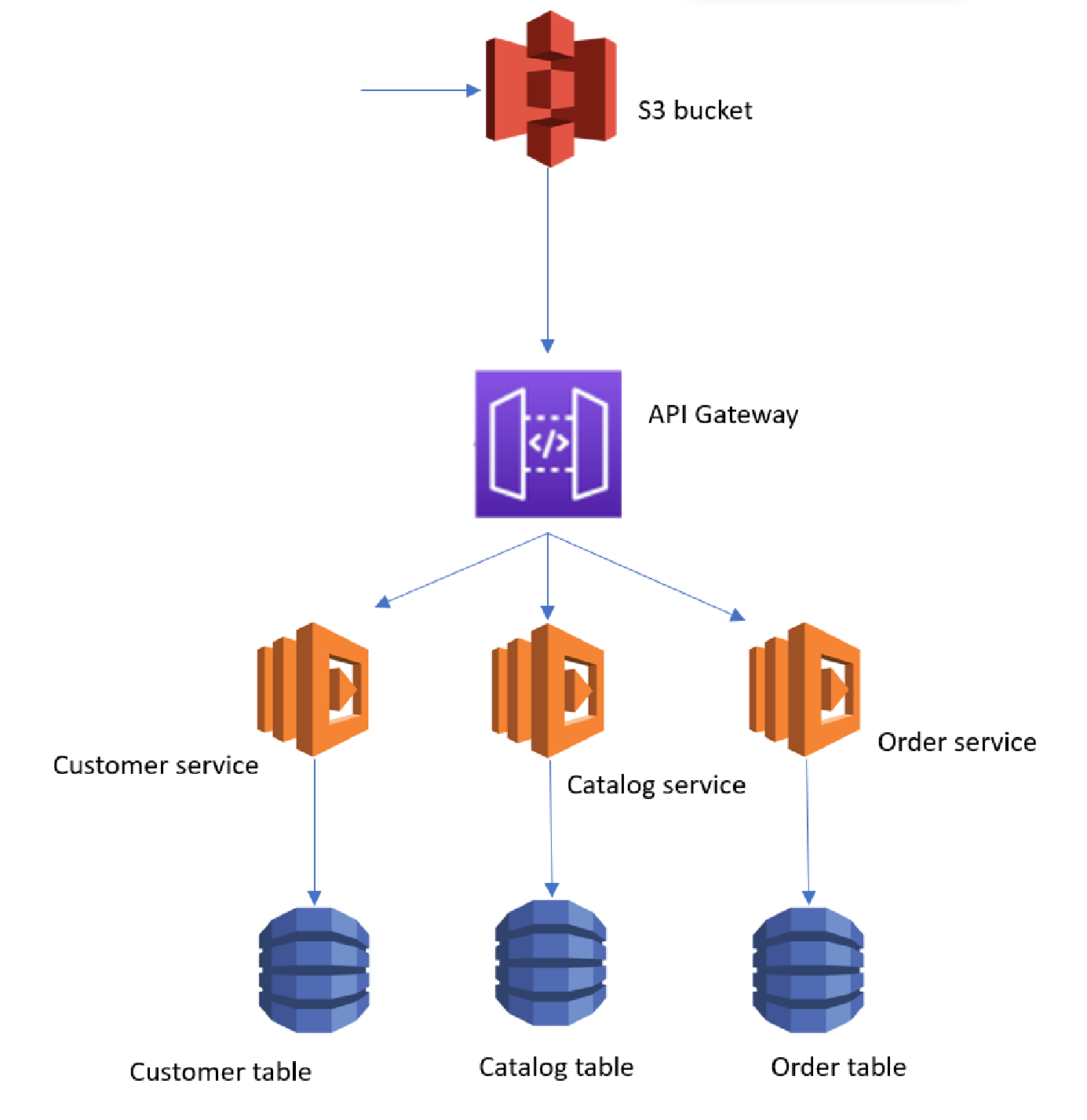}
			\caption[Initial application architecture]{Initial application architecture}
			\label{fig:initial architecture}
		\end{figure}
		
		\section{Initial application}
		A common use case for AWS Lambda is to integrate with DynamoDB, allowing functions to perform operations on data stored in DynamoDB. Following this approach, an initial solution has been built to measure its initial performance and demonstrate the performance improvements generated as a result of the proposed best practices.\\
		The mentioned system consists of a simple e-commerce solution, compound of three services with functionality which translates in one function for each service. All the services are self-contained and individual, offering an API through of which the services send messages (using AWS API Gateway). The three services are the following:
		\begin{itemize}
			\item Catalog service: It manages the operations related with the products.
			\item Customer service: It manages everything related with customers.
			\item Order service: It involves all in relation to orders.
		\end{itemize}
		The scheme of the system will be better explained in the Figure~\ref{fig:initial architecture}. The performance of this initial solution can be seen in Table \ref{tab:initial_solution_performance} with an average latency of 512 ms when lambdas are warmed and over 16307 ms on cold starts. It is important to highlight that the test lambda functions are mainly based on interaction with DynamoDB and are not computationally intensive.
        \begin{table}[h]
            \centering
            \begin{tabular}{|c|c|c|c|} \hline
                \multicolumn{4}{|c|}{Initial Solution (warmed functions)} \\ \hline
                functions / workload & 1000 users  & 5000 users & 10000 users \\ \hline
                customer lambda & 482 ms & 383 ms & 390 ms \\ \hline
                catalog lambda & 564 ms & 487 ms & 495 ms \\ \hline
                order lambda & 657 ms & 589 ms & 581 ms \\ \hline
            \end{tabular}
            \caption{Performance measurement of initial solution}
            \label{tab:initial_solution_performance}
        \end{table}

        \section{Performance tests}
        In line with the main objective of the study, load tests will be conducted for each of the implemented improvements. The objectives of these tests is quantify the improvement offered by each of the implemented techniques or approaches and compare the latencies with the initial unoptimized system. \\
        Due to that, the tests have a duration of 30 minutes in which the tested system is exposed to different constant workloads. Each test is repeated a total of 3 times, with the result being the average value of all repetitions. To evaluate cold start performance, functions were invoked manually from the AWS console, and the experienced times were collected. A test is compound of 3 requests, one for each function of the application. The results will be measured and displayed in milliseconds (ms).
		
        \section{Best practices and techniques} 
		Below are the proposed techniques and approaches to reduce the cold start of Java functions and improve their overall performance. Along with their measurements in terms of performance tests. The proposed practices will be compared in the \ref{Discussion}.Discussion section.
  
		\subsection{Better suited configuration}
\begin{figure}[h]
			\centering
			\includegraphics[width=1\linewidth]{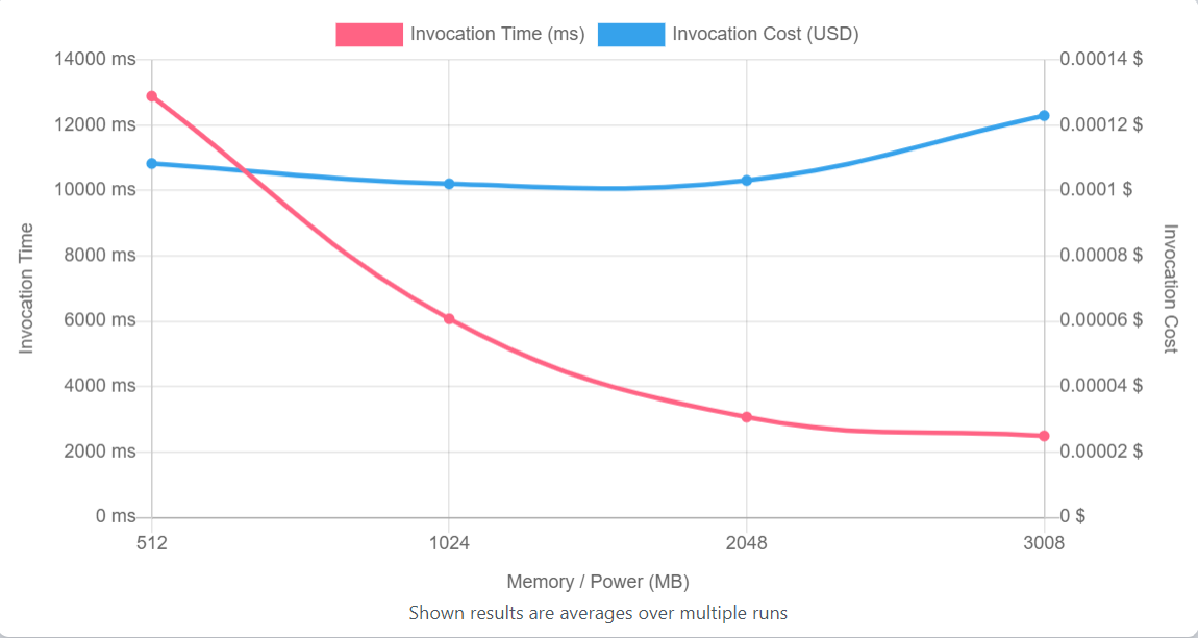}
			\caption[Power tuner with Catalog lambda results]{Power tuner with Catalog lambda results}
			\label{fig:catalog_lambda_powertuned}
		\end{figure}

        Increasing the allocated memory for a function is probably the most well-known way to reduce cold starts and also the simplest. The allocation of memory to a function is related to the amount of CPU assigned to that function. A greater amount of CPU generally makes the initialization of the function faster. However, it's not always the case that more memory equals better performance. It's also necessary to consider that higher memory allocation will result in higher costs, but if the computing time is reduced due to increased performance, the overall cost will decrease as well. To properly balance the memory configuration of a Lambda function, the Lambda Power Tuner solution \parencite{AWSProfFunc} can be of great help to profile the function, displaying graphs such as the one shown in Figure \ref{fig:catalog_lambda_powertuned} for Catalog lambda of our solution.\\
        In our case, chose the right memory value for the functions configuration suppose an improvement of about 70\% in terms of cold start going from an average of 16307 ms to 4867 ms. In addition to a noteworthy 46\% when the functions are warmed passing from an average of 512 ms to 216 ms as can be seen in Table \ref{tab:conf_performance}. For the rest of the versions that do not have an appropriate memory configuration, the default value of 512 MB is used.
        \begin{table}[h]
            \centering
            \begin{tabular}{|c|c|c|c|} \hline
                \multicolumn{4}{|c|}{Configuration optimized solution (warmed functions)} \\ \hline
                functions / workload & 1000 users & 5000 users & 10000 users \\ \hline
                customer lambda & 305 ms & 244 ms & 257 ms \\ \hline
                catalog lambda & 321 ms & 249 ms & 264 ms \\ \hline
                order lambda & 312 ms & 258 ms & 277 ms \\ \hline
            \end{tabular}
            \caption{Performance measurement of configuration optimized solution}
            \label{tab:conf_performance}
        \end{table}
        
        \subsection{Snapstart}
        Snapstart \parencite{AWSsnapstart} is a feature offered by AWS to mitigate the cold start of Java functions. It involves initializing the function at the moment a version of it is created. Lambda takes a snapshot of the initialized lambda's runtime environment, encrypts and caches it. When initializing new functions, it does so from the snapshot rather than starting from scratch, thus improving performance.\\
        A notable drawback of this feature is that it does not support many other features offered by AWS Lambda, such as custom runtimes, the arm64 architecture, Amazon EFS, and other managed runtimes that are not Java 11 or Java 17.\\
        In this case, enable Snapstart on our functions allow an enhancement of 16\% at the cold start (going from an average of 16307 ms to 13736 ms) and an improvement of 21\% attending requests when the functions are warmed making a difference of over 100 ms in each request answered (Table \ref{tab:snapstart_performance}).
        \begin{table}[h]
            \centering
            \begin{tabular}{|c|c|c|c|} \hline
                \multicolumn{4}{|c|}{Snapstart (warmed functions)} \\ \hline
                functions / workload & 1000 users & 5000 users & 10000 users \\ \hline
                customer lambda & 427 ms & 306 ms & 364 ms \\ \hline
                catalog lambda & 488 ms & 271 ms & 410 ms \\ \hline
                order lambda & 560 ms & 282 ms & 525 ms \\ \hline
            \end{tabular}
            \caption{Performance measurement of Snapstart solution}
            \label{tab:snapstart_performance}
        \end{table}
        
        \subsection{Arm64 architecture}
        The instruction set architecture within AWS Lambda, either x86\_64 or arm64, determines the processor that the function will use. AWS's official documentation suggests that the arm64 architecture \parencite{AWSarm64} with its Graviton processor can offer significantly better performance than the x86\_64 architecture. \\
        Regarding our solution, the impact of using arm64 on the latencies was positive, getting over a 14\% of improvement on cold starts and a 13\% when the lambdas are warmed. Reducing the average latency by over 2000 ms and approximately 70 ms, respectively. The warmed lambda functions data can be seen in Table \ref{tab:arm64_performance}.
        \begin{table}[h]
            \centering
            \begin{tabular}{|c|c|c|c|} \hline
                \multicolumn{4}{|c|}{arm64 (warmed functions)} \\ \hline
                functions / workload & 1000 users & 5000 users & 10000 users \\ \hline
                customer lambda & 374 ms & 376 ms & 355 ms \\ \hline
                catalog lambda & 455 ms & 440 ms & 418 ms \\ \hline
                order lambda & 542 ms & 543 ms & 506 ms \\ \hline
            \end{tabular}
            \caption{Performance measurement of arm64 solution}
            \label{tab:arm64_performance}
        \end{table}
        
        \subsection{AWS SDK v2 for Java}
        The AWS SDK for Java 2.x \parencite{AWSsdkv2} is the second version of the SDK provided by AWS for Java. This version is designed with a modular approach, allowing you to include only the service-specific JARs you need, reducing the size of your artifacts. That's a key point of reducing the cold start, loading less number of classes when initialize the function. Furthermore, it boasts several other performance improvements such as asynchronous nature, connection management, concurrency support, resource management, and so forth. Due to that, in terms of performance its use is highly recommended. \\
        For us, migrate the lambda functions from sdk v1 to sdk v2 suppose an enhancement of about 40\% in both cold start and warmed functions. On first going from 16307 ms to 9926 ms and on second one passing from 512 ms to 298 ms of latency as is visible in Table \ref{tab:sdkv2_performance}.
        \begin{table}[h]
            \centering
            \begin{tabular}{|c|c|c|c|} \hline
                \multicolumn{4}{|c|}{sdk v2 (warmed functions)} \\ \hline
                functions / workload & 1000 users & 5000 users & 10000 users \\ \hline
                customer lambda & 307 ms & 277 ms & 267 ms \\ \hline
                catalog lambda & 300 ms & 290 ms & 279 ms \\ \hline
                order lambda & 340 ms & 317 ms & 305 ms \\ \hline
            \end{tabular}
            \caption{Performance measurement of sdkv2 solution}
            \label{tab:sdkv2_performance}
        \end{table}

        \subsection{GraalVM}
        GraalVM \parencite{graalvm} is a platform that enables ahead-of-time compilation of Java code, leveraging the Native Image tool to produce a standalone binary. As a result, this binary offers significantly better performance compared to traditional Java code that runs using the JVM and is compiled at runtime. The binary is produced for the OS of the environment the code is built. It's important to note that AWS Lambda is only compatible with GraalVM through the use of custom runtimes. Other disadvantage is that the ahead-of-time compilation sometimes would be tricky to make it work properly adding even the necessity of dealing with low level issues caused by AOT compilation. \\
        For us, create a native image of our code and uses it generates a great improvement of 83\% on cold start, going form 16307 ms to 2800 ms. When the lambdas are warmed, it goes from and average of 512 ms to 231 ms which means a 55\% of enhancement, visible in Table \ref{tab:graalvm_performance}.
        \begin{table}[h]
            \centering
            \begin{tabular}{|c|c|c|c|} \hline
                \multicolumn{4}{|c|}{GraalVM (warmed functions)} \\ \hline
                functions / workload & 1000 users & 5000 users & 10000 users \\ \hline
                customer lambda & 269 ms & 193 ms & 231 ms \\ \hline
                catalog lambda & 272 ms & 194 ms & 229 ms \\ \hline
                order lambda & 265 ms & 196 ms & 230 ms \\ \hline
            \end{tabular}
            \caption{Performance measurement of GraalVM solution}
            \label{tab:graalvm_performance}
        \end{table}
        
        \subsection{Environment variable: JAVA\_TOOL\_OPTIONS}
        AWS Lambda provides some customization options \parencite{AWSsettings} for the managed Java runtime, which can enhance overall function performance. This is achieved through the use of the JAVA\_TOOL\_OPTIONS environment variable, which allows access to JVM features such as tiered compilation or garbage collector behavior among others. In this case we use only tiered compilation, fixing the level to 1, enabling the C1 compiler which means optimized code for fast start-up time. The exact value for the mentioned key is "-XX:+TieredCompilation -XX:TieredStopAtLevel=1". Java 17 managed runtime is default set to level 1 for AWS Lambda, while in Java 11, you need to set it manually. Level two can be used for an overall better performance, not only focus on the start time.
        In our test system, enabling that option translates to a 40\% improvement in cold start times and a 31\% improvement in response times for the system with warm Lambdas (Table \ref{tab:java_options_performance}).
        \begin{table}[h]
            \centering
            \begin{tabular}{|c|c|c|c|} \hline
                \multicolumn{4}{|c|}{JAVA\_TOOL\_OPTIONS (warmed functions)} \\ \hline
                functions / workload & 1000 users & 5000 users & 10000 users \\ \hline
                customer lambda & 359 ms & 311 ms & 327 ms \\ \hline
                catalog lambda & 368 ms & 336 ms & 350 ms \\ \hline
                order lambda & 402 ms & 369 ms & 365 ms \\ \hline
            \end{tabular}
            \caption{Performance measurement of java\_options solution}
            \label{tab:java_options_performance}
        \end{table}

        \subsection{Others}\label{Others}
        In addition to the previous techniques and approaches we propose and have tested, there are many other best practices \parencite{AWSpractices,StefanoBuliani,OToole2020}. Whether they are more dependent on the use case or more general, below, we mention some of them:
        \begin{itemize}
			\item Reduce the size of the bundle to enhance the initialization phase. As less classes to load as fast the function will initialize.
			\item Initialize the connections of third party out of the handler and same the static resources on /tmp to reuse it from one execution to another.
			\item A framework, if needed, should be as lighter as possible. Due to that we suggest replacing spring with other smaller frameworks like Vert.x
            \item Try to avoid reflection, that feature makes some JVM optimizations impossible to be done as it works with dynamic types.
            \item Initialize all necessary dependencies and classes during initialization time.
		\end{itemize}
        These approaches, combined with some of the previous ones we have tested, for sure will greatly reduce the impact of cold starts and increase the performance of Java functions by a really significant percentage.

        \section{Combinations}
        After defining the proposed techniques and quantifying the effects of their application, it becomes essential to attempt to combine them with the intention of detecting synergies, incompatibilities, and understanding the improvements that can be achieved by applying several of these. With this intention, and due to the unavailability of Snapstart in combination with the Arm64 architecture, the following combinations are proposed:
        \begin{itemize}
            \item Best memory configuration + SDK v2 + JAVA\_TOOL\_OPTIONS environment variable + Snapstart
            \item Best memory configuration + SDK v2 + JAVA\_TOOL\_OPTIONS environment variable + Arm64 architecture
        \end{itemize}
        Two new prototypes have been created, optimized with the previous optimization combinations, and their performance has been measured. The results are compared with each other and independently with the GraalVM prototype.\\
        \begin{figure}[h] 
			\centering
			\includegraphics[width=1\linewidth]{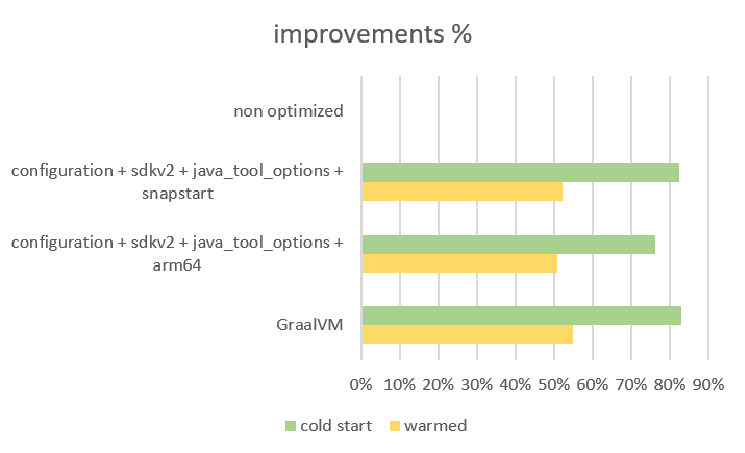}
			\caption[Cold start enhancement by approach \%]{Percentage of improvement by combination \%}
			\label{fig:combination_performance}
		\end{figure}
        In Figure \ref{fig:combination_performance}, you can observe the percentage of improvement compared to the performance offered by the initial system, both for the cold start of functions and for the overall performance.
        
        \section{Discussion}\label{Discussion}
\begin{figure}[h] 
			\centering
			\includegraphics[width=1\linewidth]{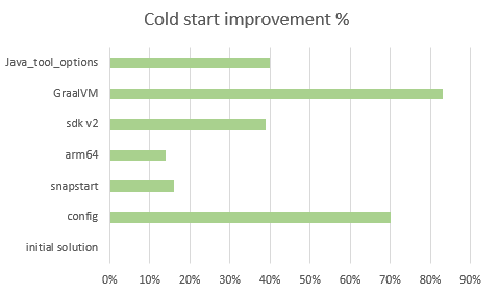}
			\caption[Cold start enhancement by approach \%]{Cold start enhancement by approach \%}
			\label{fig:coldstart_technique_separately}
		\end{figure}

\begin{figure}[h] 
			\centering
			\includegraphics[width=1\linewidth]{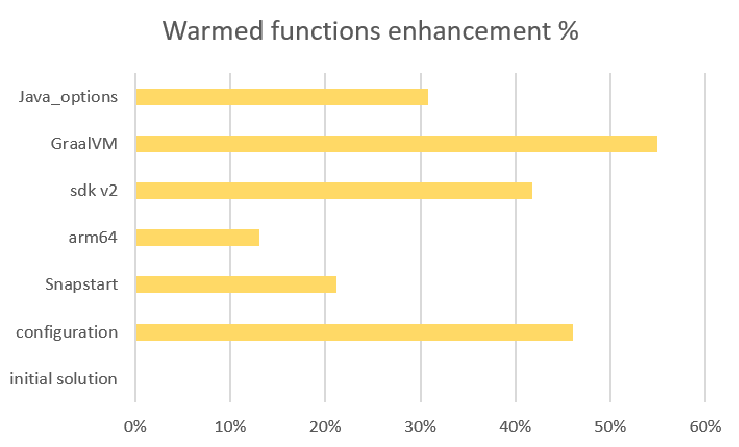}
			\caption[Warmed function enhancement by approach \%]{Warmed function enhancement by approach \%}
			\label{fig:warmed_technique_separately}
		\end{figure}

        In Figure \ref{fig:coldstart_technique_separately} and Figure \ref{fig:warmed_technique_separately}, you can see a summary of the cold start improvement and warm Lambdas, respectively. Next, we will analyze each approach and the possibilities they offer. \\
        Starting with an appropriate configuration for each function, in our opinion, is the minimum optimization that any function should have, both to improve performance and to balance cost. In addition to offering a significant improvement, it is perfectly compatible with all the other techniques mentioned in this article. \\
        Continuing with the use of Snapstart, this technique focuses on mitigating cold starts. In our experience, this technique could not produce as significant an improvement as to justify the incompatibilities it generates, including the lack of support for the arm64 architecture or custom runtimes, along with other important features such as the capability to use EFS or to attach the lambda to a VPC. \\
        The improvement offered by using the Arm64 architecture seems to depend on the specific use case in which it is employed. For example, it can be particularly beneficial for compute-intensive applications such as high-performance computing. This could explain why the improvement in our case is not too noticeable. Nevertheless, its use is highly recommended as it can be combined with other approaches discussed, and its adoption is on the rise. The major drawback is the incompatibility with some third-party technologies or dependencies related to this architecture.\\
        The use of AWS's SDK v2 for Java applications is also practically mandatory, as long as it supports all the services and the majority of the dependencies integrated into the function. Thanks to its improvements and modular approach, this second version of the SDK significantly increases performance. Moreover, it is perfectly mixable with the rest of the techniques without adding notable limitations. Combined with version 1, if any unsupported features are needed, they can be utilized as required. \\
        As expected, leveraging the benefits of ahead-of-time compilation generates a significant performance boost both in cold starts and when functions are already warm. However, the main drawback is the complexity that this approach adds. It's common to encounter runtime issues that may require the developer to deal with low-level aspects, and any modifications can become challenging. In principle, it is perfectly combinable with the rest of the techniques that support custom runtimes. \\ 
        The possibilities offered by the JAVA\_TOOL\_OPTIONS environment variable are quite extensive, allowing for significant Java VM configuration. In our case, we only tested tiered compilation, but there are other possibilities such as configuring garbage collector behavior \parencite{gcControlling}. In any case, it's worth trying to use this variable and integrating it into Java functions whenever the JVM is used because it doesn't impose any limitations, and the improvement is significant.\\
        It's also important to highlight that investing efforts in refactoring the function's code and applying the details mentioned in subsection \ref{Others} can achieve an even greater performance improvement than the techniques discussed. This work is entirely dependent on the use case and requires trial and error as well as skill. \\
        Regarding the simultaneous application of techniques, for our combinations, we recommend the use of either of the two, so you should choose the one that provides the most benefit, either in terms of performance or greater adaptability to the specific use case.

        \section{Related work}
        As far as we know, and up to this date, there aren't too many articles seeking to enhance the Java approach with AWS Lambda in the same manner as we do. Nonetheless, there are some similar studies. In \parencite{StefanoBuliani} Buliani addresses part of the performance issue and proposes and applies a series of techniques incrementally, analyzing the overall improvement step by step. \parencite{performanceWu} investigates the performance impact of cold start and containerization for Java-based FaaS functions and compares it to an alternative using Python. \parencite{Bardsley} takes a similar approach, suggesting a variety of ways to enhance the performance of AWS Lambda but without integrating it with Java. In \parencite{Dowd2020} Dowd investigates the overhead and cold start for functions written in Java, attempting to determine the extent to which the findings depend on the programming language. \\
        There are also other studies that have a certain relationship with ours, either because they explore some of the techniques we propose or because they are related with the performance of AWS Lambda. \parencite{Jackson, Hosseini2019} analyze the performance of the various managed runtimes provided by AWS for executing code in different languages on Lambda functions without optimizing it. In \parencite{Hussachai} Hussachai et al demonstrates how optimizing the Java artifact results in superior performance during the execution of the Lambda function. In \parencite{gcControlling} evaluate the use of a garbage collector control strategy concerning performance improvement, that is an interesting strategy that we didn't take in considerationin our study. Related with GraalVM, \parencite{Mosquera2022, graalvmsipek} incorporate this tool with the aim of leveraging the advantage it offers in terms of performance.
  
		\section{Conclusion} 
        Performance is an aspect that should always be taken into account when creating Lambda functions, especially if you plan to combine them with Java. Once the performance improvements offered by these approaches for using Java with AWS Lambda have been quantified. It is evident that Java can be used competitively with AWS Lambda without the need to resort to runtimes like Node.js or Python (which usually report better performance). \\
        Having tested various combinations of proposed strategies and demonstrated that the performance is similar to what can be achieved with the use of GraalVM, we recommend avoiding its usage. We consider it essential to profile the Java functions to select the optimal memory configuration and using AWS SDK v2. From there, optionally, the next step is to the rest of the techniques to achieve one of the proposed combinations.
        Additionally, following the general best practices such as initializing connections outside the handler or reduce the bundle size, among others, also contribute to optimizing Java functions. \\ 
        The optimization strategies we propose are not the only ones available, there are many other techniques that can similarly yield excellent results.\\
        It is important to highlight that the process of optimizing a function is linear in terms of difficulty. In other words, it's relatively easy to obtain a significant improvement, but if you want to squeeze out the maximum performance, the process requires trial and error and greater effort. It's also important to highlight that the percentage of improvement offered by each approach depends on the specific use case in which they are applied. This percentage can vary, either increasing or decreasing.
		\printbibliography
	\end{document}